\documentclass[11pt,aps,superscriptaddress]{article}

\usepackage{graphicx,graphics,subfigure}
\usepackage{amsfonts,amsmath,amssymb,mathrsfs,bm,amsthm,nccmath}

\usepackage{array}
\usepackage{tabu}
\usepackage{dcolumn}
\usepackage{float}
\usepackage{url}
\usepackage{ragged2e}
\usepackage{mathtools} 
\usepackage{multirow} 
\usepackage{slashed}
\usepackage[usenames,dvipsnames]{color}
\usepackage{soul}
\usepackage{url}
\usepackage[colorlinks = true,
            linkcolor = blue,
            urlcolor  = blue,
            citecolor = blue,
            anchorcolor = blue,linktocpage]{hyperref}
\usepackage{physics}
\usepackage{makecell}
\usepackage{units}
\usepackage{upgreek}

\usepackage[total={6.5in,9in},top=1in,headsep=0.1in,headheight=1in]{geometry}
\usepackage[utf8]{inputenc}
\usepackage{charter}
\usepackage{fullpage}
\usepackage{hyperref}

\def\bea{\begin{eqnarray}}
\def\eea{\end{eqnarray}}

\definecolor{Zpurple}{RGB}{119, 50, 168}

\definecolor{cardinal}{RGB}{140, 21, 21}

\newcommand\snowmass{
\begin{center}
  \rule[-0.2in]{\hsize}{0.01in}\\
  \rule{\hsize}{0.01in}\\
  \vskip 0.1in
  Submitted to the Proceedings of the US Community Study\\
  on the Future of Particle Physics (Snowmass 2021)\\
  \rule{\hsize}{0.01in}\\
  \rule[+0.2in]{\hsize}{0.01in}\\[-2em]
\end{center}
}

\usepackage[firstpage=true]{background}
\backgroundsetup{contents={\parbox{6.5in}{\snowmass}}, scale=1,placement=top,opacity=1,color=black,position={3.25in,1.2in}}

\usepackage{fancyhdr}
\fancypagestyle{plain}{%
  \fancyhf{}%
  \fancyhead[C]{}
  \fancyfoot[C]{\thepage}
}

\usepackage{authblk}

\usepackage{lipsum}

\title{Snowmass2021 - White Paper\\
Collider Physics Opportunities of Extended Warped Extra-Dimensional Models}
\date{}

\author[1]{Kaustubh Agashe}
\affil[1]{Maryland Center for Fundamental Physics, Department of Physics, University of Maryland, College Park, MD 20742, USA}

\author[2]{Jack H. Collins}
\affil[2]{SLAC National Accelerator Laboratory, 2575 Sand Hill Road, Menlo Park, CA 94025, USA}

\author[3]{Peizhi Du}
\affil[3]{C.N. Yang Institute for Theoretical Physics, Stony Brook University, Stony Brook, NY, 11794, USA}

\author[4]{Majid Ekhterachian}
\affil[4]{Theoretical Particle Physics Laboratory (LPTP),
Institute of Physics, EPFL, Lausanne, Switzerland}

\author[5,6]{Sungwoo Hong}
\affil[5]{Department of Physics, The University of Chicago, 5720 S Ellis Ave, Chicago, IL 60637 , USA}
\affil[6]{Argonne National Laboratory, 9700 S. Cass Avenue, Lemont, IL 60439, USA}

\author[7]{Doojin Kim}
\affil[7]{Mitchell Institute for Fundamental Physics and Astronomy,
Department of Physics and Astronomy, Texas A\&M University, College Station, TX 77845, USA}

\author[8]{Rashmish K. Mishra}
\affil[8]{Harvard University, 17 Oxford Street, Cambridge, MA, 02138, USA}

\author[1]{Deepak Sathyan}

\begin{document}
\maketitle

\begin{abstract}
While the warped extra dimensional models provide an attractive solution to both the gauge and the flavor hierarchy problems, the mass scale of new particles predicted by the minimal models would be beyond the reach of the LHC. Models of extended warped extra dimensions have been proposed to evade these issues and their collider implications have been investigated for the last decade. This white paper summarizes the recent developments in the context of collider phenomenology. The strategies and lessons are broad, and provide a template to extend the experimental program, to cover a wider class of signals in other new physics scenarios as well.
\end{abstract}

\clearpage

\tableofcontents

\section{Executive Summary}
Although the monumental discovery of the Higgs particle puts the final piece of the Standard Model (SM), the stability of its mass against high-scale physics still remains as an open question that the SM cannot answer. The Randall-Sundrum model with two branes~\cite{Randall:1999ee,Randall:1999vf} is a promising framework to address this issue, and the warped extra-dimensional models~\cite{Sundrum:2005jf,Csaki:2005vy,Davoudiasl:2009cd,Ponton:2012bi}, where even the SM gauge and matter fields propagate in the bulk, additionally provide an attractive solution to the flavor hierarchy problem. These models predict rich phenomenology at the Large Hadron Collider (LHC), but no conclusive signals have been observed yet; in fact, flavor/CP violation constraints (without extra symmetries) already suggest that the mass scale of relevant new particles be of $\mathcal{O}(10)$~TeV~\cite{Csaki:2008zd,Blanke:2008zb,Bauer:2009cf,KerenZur:2012fr} even before the LHC was turned on.  

In light of this situation, models of extended warped extra dimensions~\cite{Agashe:2016rle} have been proposed to avoid these potential issues while not only keeping the virtue of solving gauge/flavor hierarchy problem but letting new particles [e.g., Kaluza-Klein (KK) excitations] accessible at the LHC. 
In this white paper, we discuss the idea of models of extended warped extra dimension and their phenomenological implications on energy-frontier collider phenomenology. The signal topologies and the kinematic regimes in which they are produced, are novel and challenging for detection. The developed collider stragies are applicable to other BSM scenarios where such signals are relevant, and in general they allow the experimental programs to have sensitivity to broad class of signals.

\section{Extended Warped Model}
\begin{figure}[t]
\centering
\includegraphics[width=0.6 \linewidth]{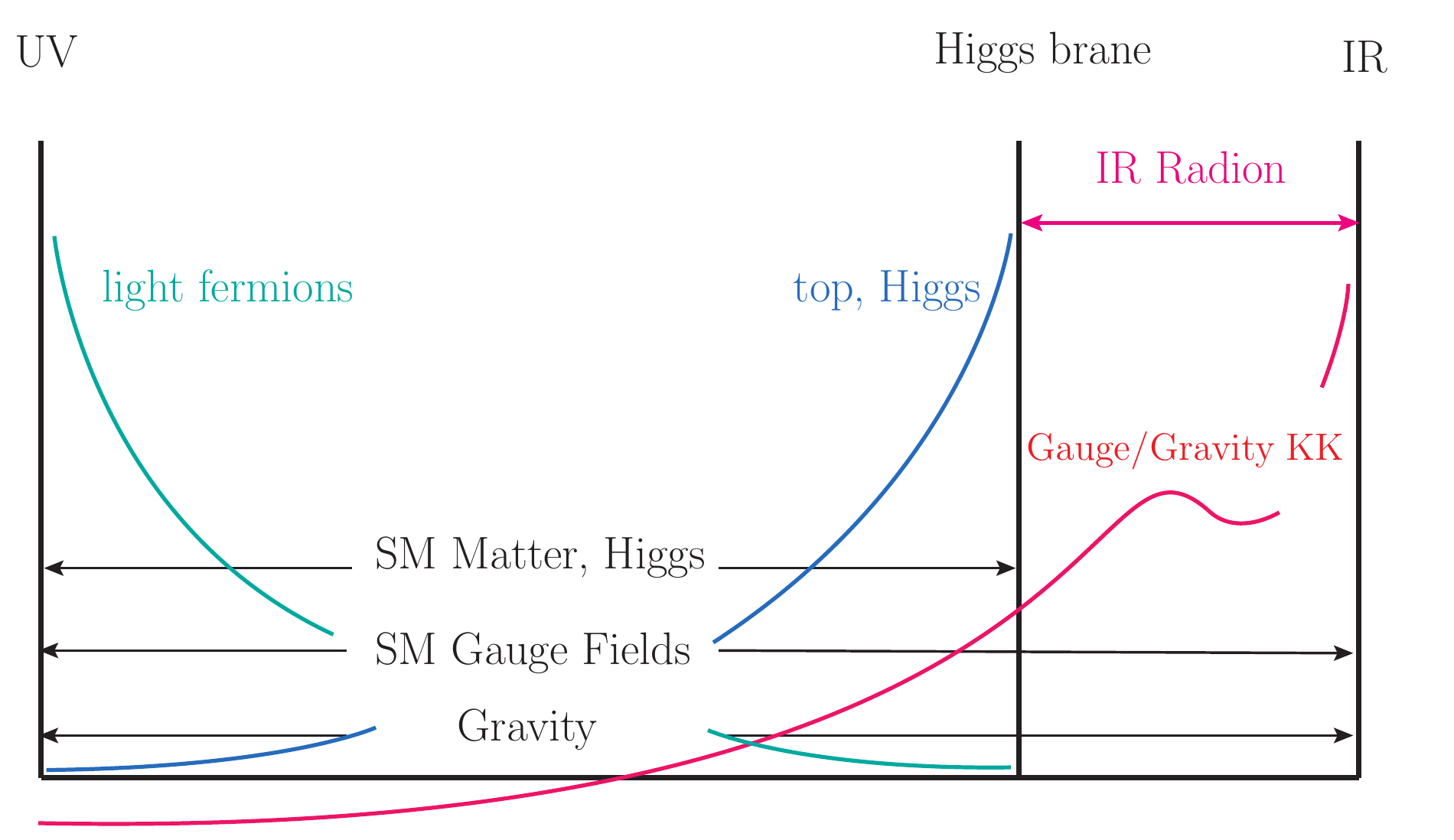}
\caption{Extended warped model with all SM gauge fields (and gravity) in the extended bulk. The colored lines show the schematic profiles of the corresponding fields.}
\label{figure:general-extended}
\end{figure}

The basic idea is to postulate an additional brane beyond the ``Higgs'' brane, where the SM Higgs field is localized, and to assume that the SM gauge fields (together with gravity) can propagate further in the extended bulk down to the additional brane (henceforth called IR brane). See also \autoref{figure:general-extended}. By contrast, the SM matter fields are confined in-between the usual UV brane and the Higgs brane, so if the Higgs brane is set to be $\mathcal{O}(10)$~TeV (at the expense of a little hierarchy issue), flavor/CP violation constraints can be avoided~\cite{Agashe:2016rle}. Now the IR brane is allowed to be of $\mathcal{O}(2-3)$~TeV without any severe tension with the existing limits, thus the LHC is capable of probing KK modes of the SM gauge fields and gravity. In addition, the usual leading decay channels of the lightest KK modes into top quark and Higgs boson are suppressed. This effect permits erstwhile subdominant channels to emerge significant.
Therefore, collider phenomenology that has received less attention becomes an important aspect in exploring extra-dimensional models at the energy-frontier facilities, throughout the upcoming decades.

\section{Collider Signatures}
In this section, we summarize various collider signatures of KK vector resonances followed by KK graviton. 

\subsection{Cascade Decays of Warped Vector Resonances} 
\begin{figure}[t]
\centering
\includegraphics[width=0.5 \linewidth]{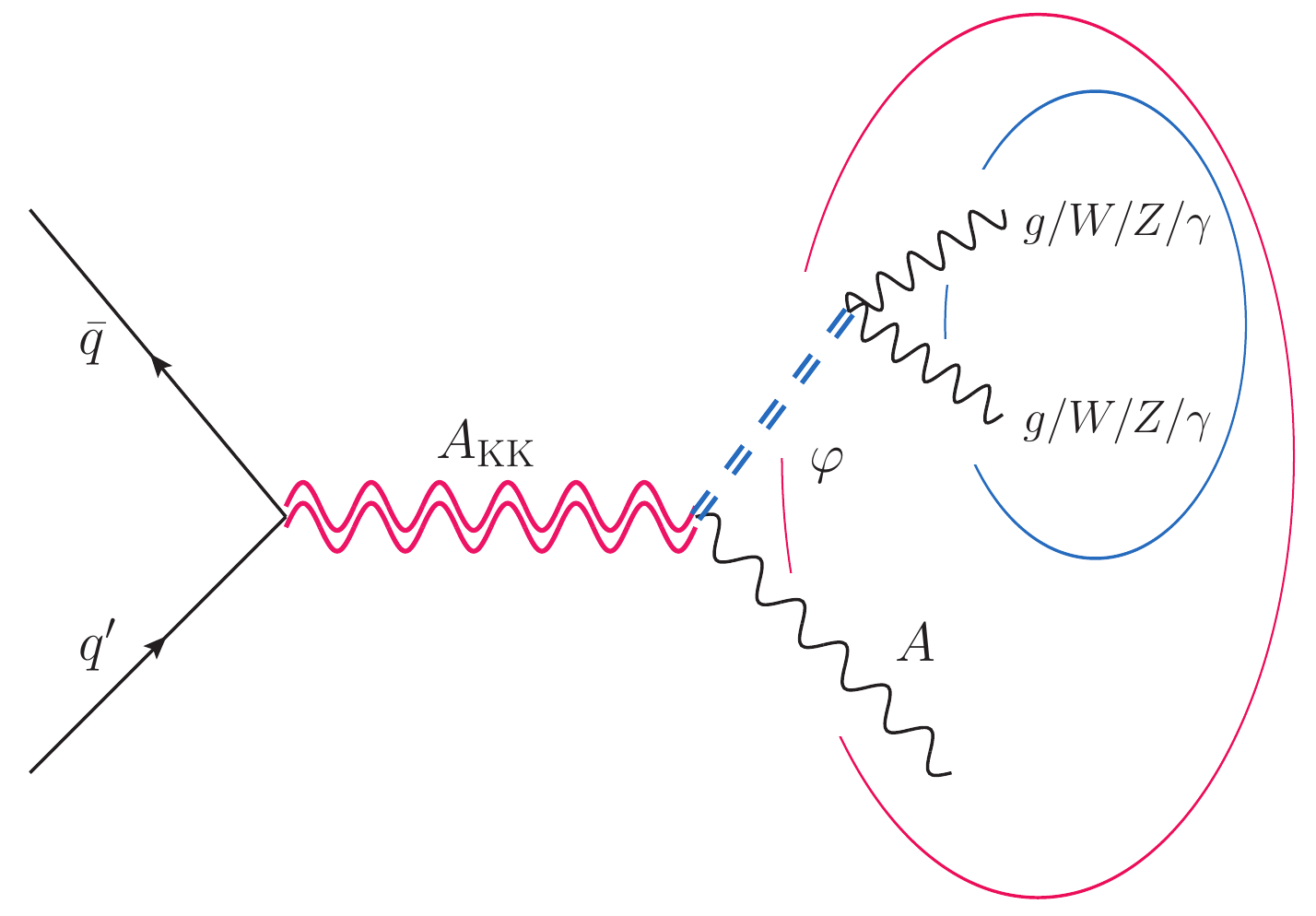}
\caption{Cascade decays of KK gauge bosons, $A_{ \rm KK }$ into a radion and the coresponding SM gauge boson,
$A = g/W/Z/\gamma$, followed by the radion decaying into a pair of {\em any} SM gauge bosons. 
}
\label{figure:KK-Radion-all}
\end{figure}

A class of well-motivated scenarios is the process where a singly produced gauge KK particle (say, $A_{\rm KK}$) decays to a corresponding SM gauge particle (say, $A$) and a radion (denoted by $\varphi$) which subsequently decays into a pair of the SM gauge particles which are possibly different from the upstream gauge particle $A$~\cite{Agashe:2016kfr} (see \autoref{figure:KK-Radion-all}):
\begin{equation}
    p p \to A_{\rm KK},~A_{\rm KK} \to A\;\varphi,~\varphi \to \gamma\gamma/ WW/ZZ/gg\,.
\end{equation}
In other words, the expected experimental signature involves a two-step cascade decay of a warped vector resonance, resulting in various combinations of three SM gauge bosons in the final state with sizable production cross-sections: for example, $\gamma gg$, $g\gamma\gamma$, $ggg$, $gV_hV_h$, and $W_l gg$, with $V_h$ and $W_l$ denoting hadronic massive SM gauge bosons and leptonic $W$, respectively. This cascade nature of the decay can make the signal identification challenging (e.g. due to combinatorial ambiguities, or due to merging of some of the final states, if sufficiently boosted). One can imagine such signals easily appearing in other contexts, making this line of investigation much more general.
A sensitivity study for those channels has been performed, reporting that they would allow for an excess of $\sim3\sigma$ to more than $\sim 10\sigma$ at the high-luminosity (HL) LHC~\cite{Agashe:2016kfr}.  Note that the radion cannot be lighter than $\sim 1$ TeV in the
scenario where {\em all} SM gauge field propagate in the extended bulk, since there are strong bounds from its di-photon decay with production from gluon fusion (both couplings are at respective full/vanilla strength). So, in this case, the two SM gauge bosons from radion decay are well-separated.

\subsubsection{Tri-EW Boson Signals}
\begin{figure}[h]
\centering
\includegraphics[width=0.49 \linewidth]{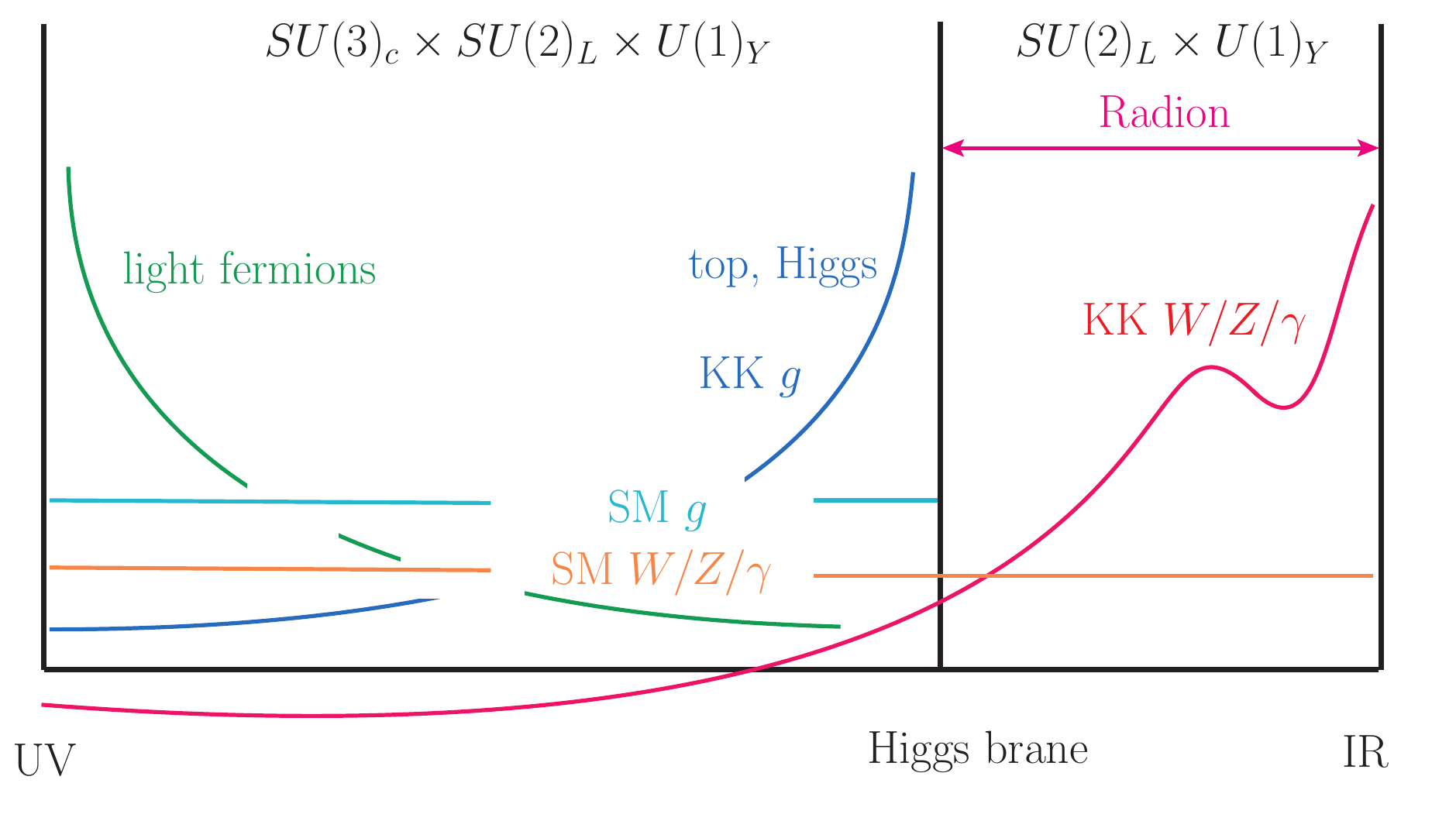}
\includegraphics[width=0.49 \linewidth]{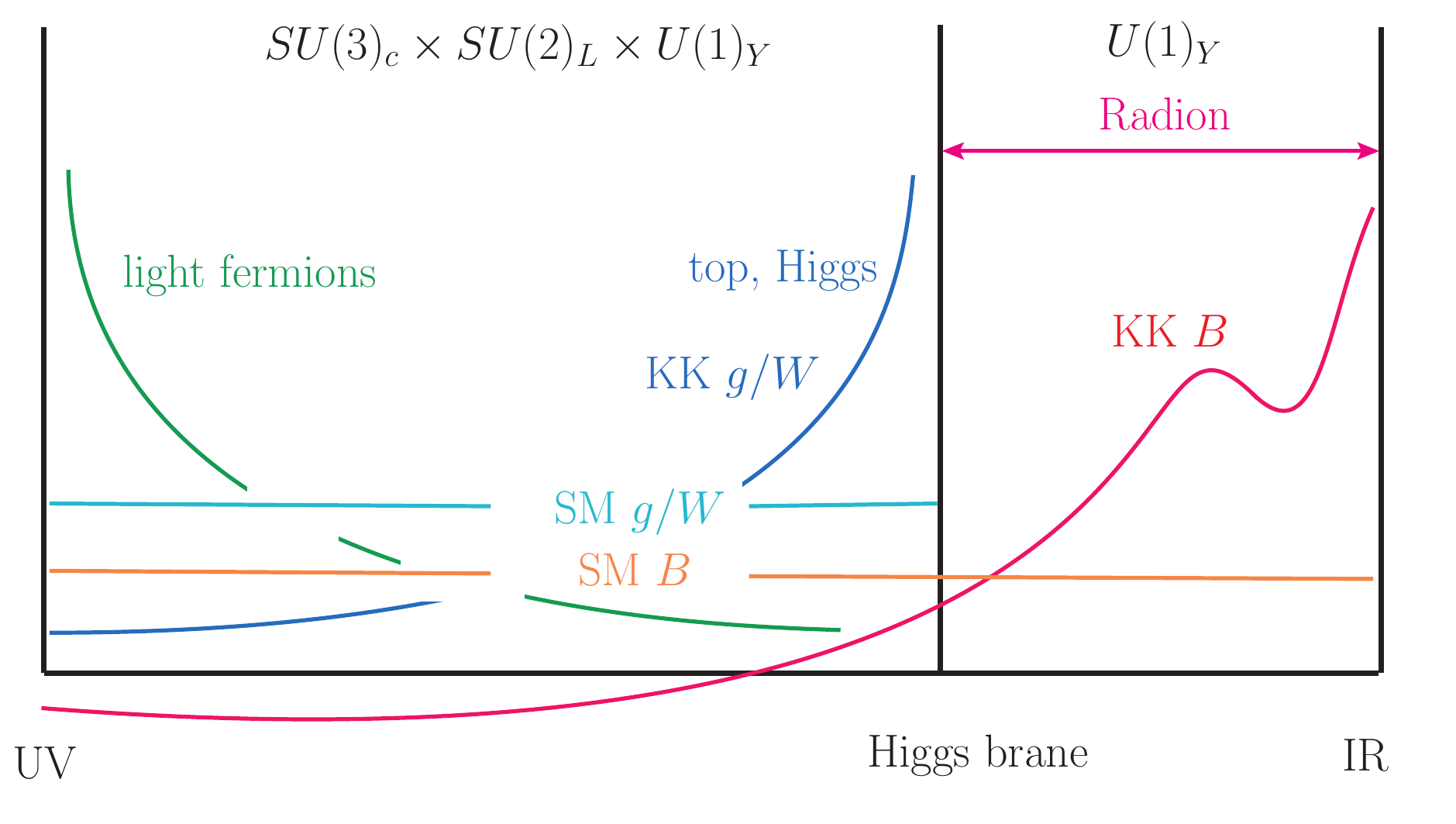}
\caption{(\textbf{Left}) Extended warped model with SM EW gauge fields in the extended bulk. (\textbf{Right}) Extended warped model with the hypercharge gauge field in the extended bulk.}
\label{figure:3Brane-EW}
\end{figure}

\begin{figure}[h]
\centering
\includegraphics[width=0.5 \linewidth]{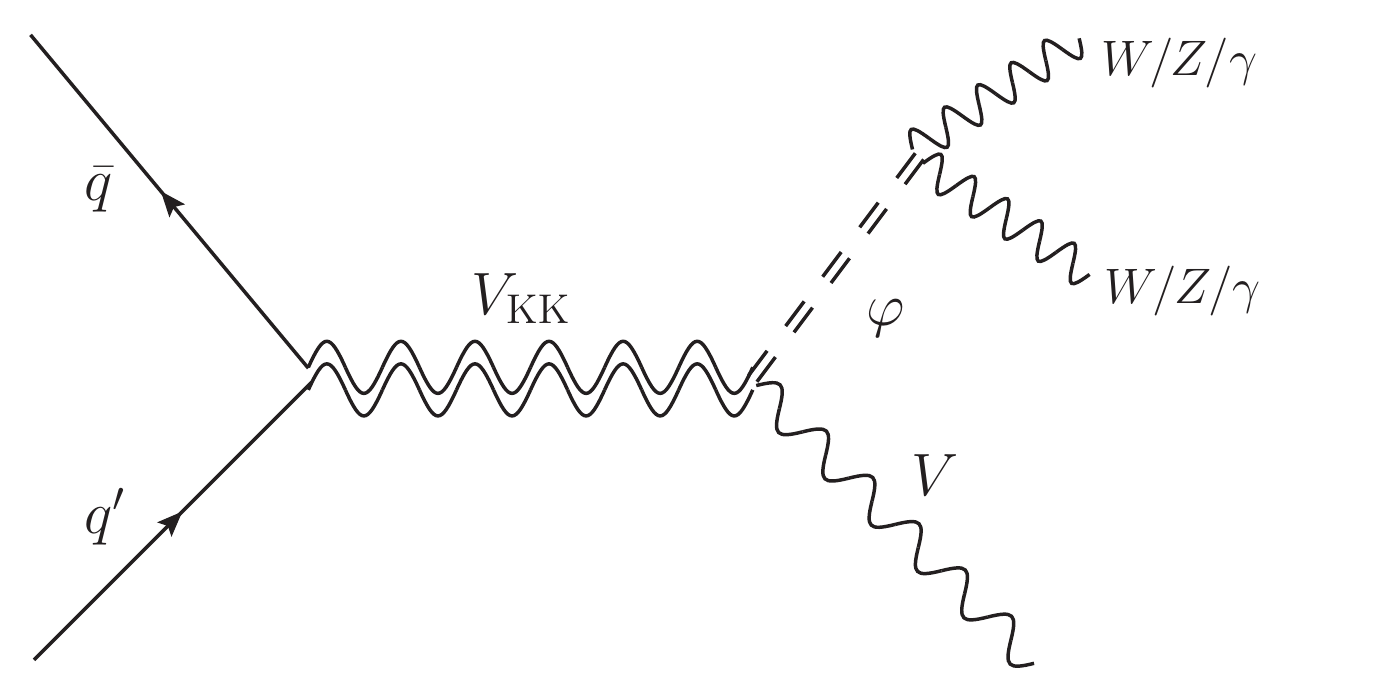}
\caption{Cascade decays of KK EW gauge bosons ($V_{ \rm KK }$) into a radion and the corresponding SM EW gauge boson, $V = W/Z/\gamma$, followed
by the radion decaying into a pair of  SM EW gauge bosons.}
\label{figure:KK-Radion-EW}
\end{figure}

An intriguing model variation under the aforementioned extended framework is to allow only electroweak (EW) gauge fields to propagate in the extended bulk while keeping the gluon field confined in-between the UV and Higgs branes~\cite{Agashe:2017wss} (see the left panel of \autoref{figure:3Brane-EW}). Under this model setup, tri-boson signals, e.g., tri-photon, tri-$W$, $W\gamma\gamma$ etc are well motivated (see \autoref{figure:KK-Radion-EW}). Depending on the mass choices, in the case of three massive gauge bosons, they can be significantly boosted so that a novel signature of three ``fat'' $W/Z$ jets can arise.  
We have performed a sensitivity study for the above tri-boson signatures, using the jet substructure techniques, and found that all channels would allow for more than $\sim 4-5\sigma$ significance even with an integrated luminosity of 300~fb$^{-1}$~\cite{Agashe:2017wss}.  

 A more specific version involves only hypercharge gauge boson having access to the extended bulk (see the right panel of \autoref{figure:3Brane-EW}), in which case KK $W$ {\em and} KK gluon are very heavy, but the KK hypercharge gauge boson is potentially within LHC reach \cite{Agashe:2017wss}.

\subsubsection{Boosted di-EW Boson Signals} 
\begin{figure}[h]
\centering
\includegraphics[width=0.6 \linewidth]{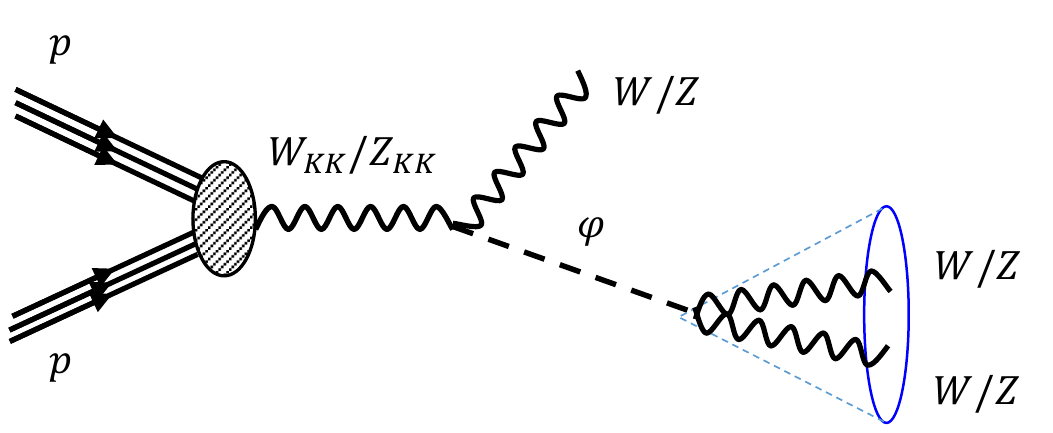}
\caption{Cascade decay of KK $W/Z$ into boosted/merged $WW/ZZ$, plus an isolated $W/Z$.
Note that this corresponds to the same Feynman diagram as in \autoref{figure:KK-Radion-EW}, but with the radion being light, thus boosted.
}
\label{figure:KKW-diW}
\end{figure}

An interesting region of parameter space for the above tri-boson signal is the region where the mass gap between the KK EW gauge and the radion is so large that the radion is significantly boosted. 
The point is that a light radion is now allowed, i.e., satisfies di-photon search bounds, since radion couplings to SM gluons, hence its production is highly suppressed, even if decays to di-photons are still present. 
As a result, its decay products, for example, two $W/Z$ are not just significantly boosted, but merged, forming a ``fat'' jet with multi-layered substructures (see \autoref{figure:KKW-diW}). This signature is not well captured by existing searches in which conventional boosted techniques are adopted, so a dedicated search strategy with a targeted jet reconstruction algorithm is needed to improve the signal sensitivity. 

\begin{figure}[h]
\centering
\includegraphics[width=0.49 \linewidth]{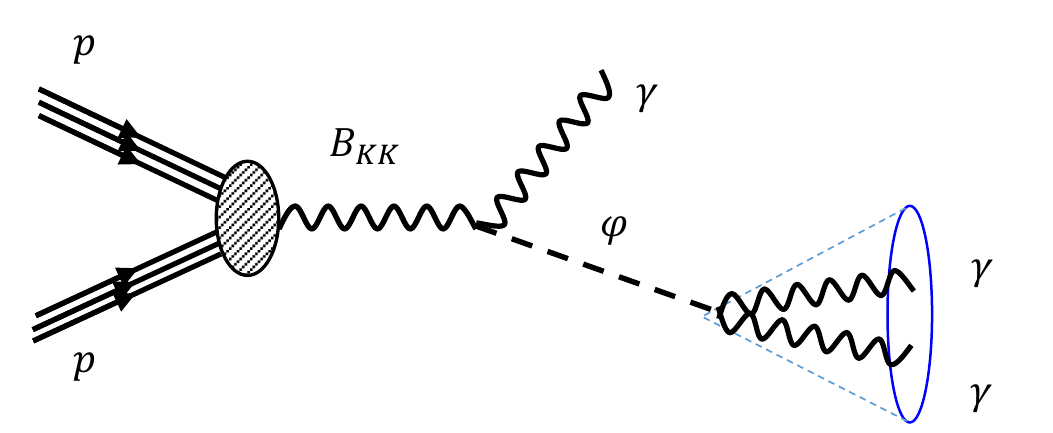}
\includegraphics[width=0.49 \linewidth]{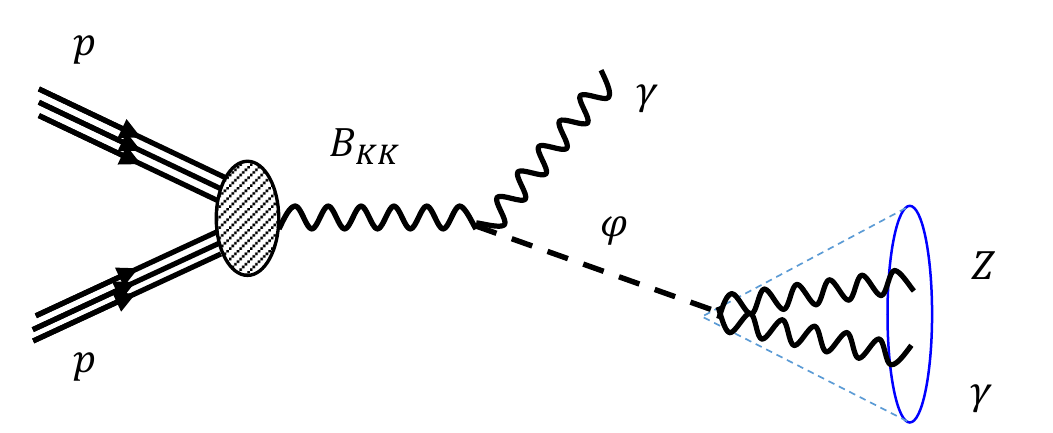}
\caption{Cascade decay of KK hypercharge gauge boson into boosted/merged di-photon (\textbf{Left}) or boosted/merged $Z+\gamma$ (\textbf{Right}), plus an isolated photon. 
Note that this corresponds to the same Feynman diagram as in \autoref{figure:KK-Radion-EW}, but with the radion being light, thus boosted.
}
\label{figure:KKphoton-phi-diphoton}
\end{figure}

We have developed respective jet substructure techniques to be sensitive to fully hadronic boosted di-boson jets  and semi-leptonic boosted di-boson jets~\cite{Agashe:2018leo}.
The application of the techniques to the fully hadronic boosted radion (i.e., $\varphi \to W_hW_h$) and the semi-leptonic boosted radion (i.e., $\varphi \to W_hW_l$) suggests that the associated searches at the LHC with an integrated luminosity of 300~fb$^{-1}$ allow us to probe regions of parameter space in the $m_{W_{\rm KK}}-m_\varphi$ plane that have been unexplored by the existing searches.
The techniques developed are generic enough to be straightforwardly applicable to similar boosted di-boson resonances in other models. Indeed, the 
fully hadronic $3 \; W$ channel along above lines has already been searched for by CMS
\cite{CMS:2021mjl} and similarly the semi-leptonic $3 \; W$ signal \cite{CMS:2022lqh}: in both studies, both light {\em and} heavy radion were considered.
One can also study hypercharge-only in extended bulk model with a {\em light} radion, 
giving 
boosted/merged di-photon (or $Z + \gamma$) from light radion decay (see \autoref{figure:KKphoton-phi-diphoton}). Presence of a light scalar is however a generic possibility, and the boosted regime of its production is quite relevant to study for the experimental programs to have a wide coverage, therefore making this line of investigation again quite broad.

\subsection{Four-jet Signals of KK Graviton}
\begin{figure}[t]
\centering
\includegraphics[width=0.9 \linewidth]{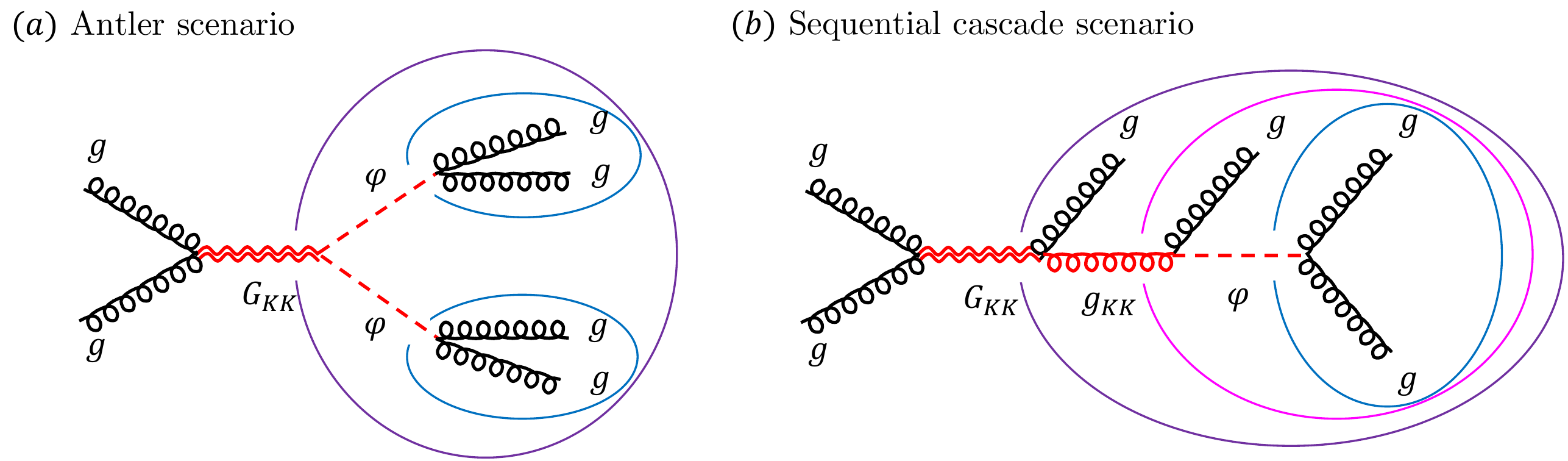}
\caption{Cascade decay of KK graviton into two radions (\textbf{Left}) or a radion, plus a KK gluon (\textbf{Right}).}
\label{figure:KK-graviton-event-topology}
\end{figure}

\begin{figure}[h]
\centering
\includegraphics[width=0.5 \linewidth]{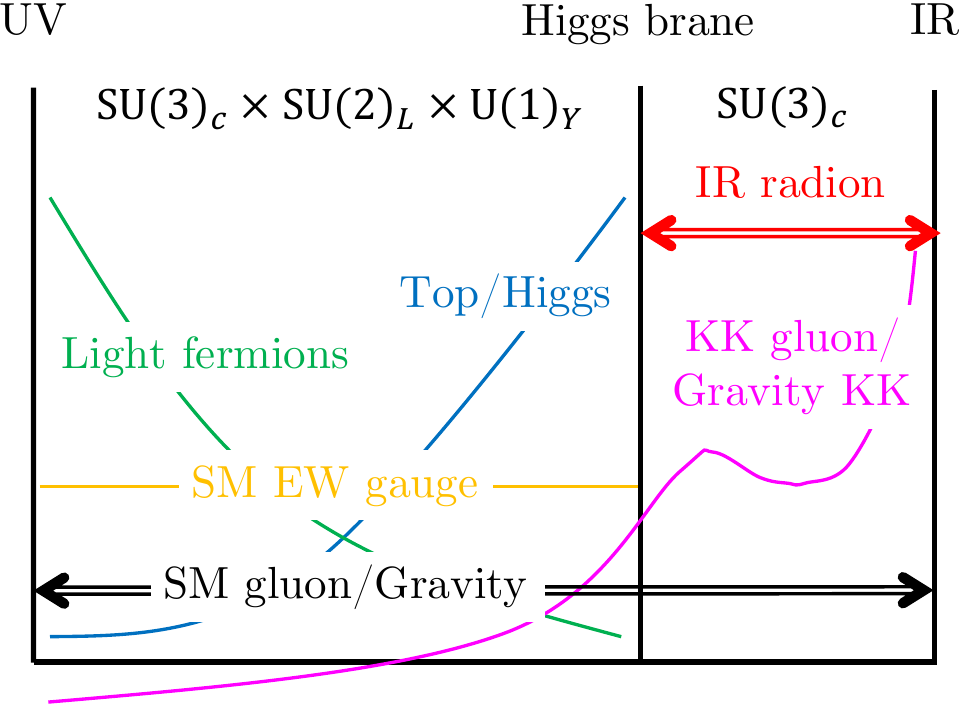}
\caption{Extended warped model with only SM gluon (and gravity) in extended bulk.}
\label{figure:QCD-gravity-extended}
\end{figure}

 As gravity always propagates in the extended bulk, its lightest KK mode can be accessible at the LHC, too. Like the KK gauge modes, its dominant decay channels into top quark and Higgs pairs in the standard framework become subdominant in the extended framework, whereas the decays to a radion pair and to a KK gauge particle and its corresponding SM gauge particle begin to ``stand out''. An interesting signature is the four-jet final state stemming from the processes, $G_{\rm KK}\to \varphi \varphi,~\varphi \to gg$ (called ``radion channel'') and $G_{\rm KK}\to gg_{\rm KK},~g_{\rm KK} \to g \varphi,~\varphi \to gg$ (called ``KK gluon channel''), with $G_{\rm KK}$ and $g_{\rm KK}$ denoting the lightest KK graviton mode and the lightest KK gluon mode, respectively (see \autoref{figure:KK-graviton-event-topology}). These channels become more significant especially in the model variation where the SM gluon field is propagating in the extended bulk while the propagation of EW fields is restricted to the Higgs brane~\cite{Agashe:2020wph} (see \autoref{figure:QCD-gravity-extended}), in part because radion decays to di-photon (golden channel) are then extremely suppressed.

Depending on the underlying mass spectrum, one channel is leading to the other, and different search strategies are motivated due to the difference of the associated event topologies. We have performed a sensitivity study with a few well-motivated benchmark sets of parameter values giving well-separated four jets, and found that the HL-LHC (3,000~fb$^{-1}$) would be sensitive to the KK graviton signals by $\sim 2.5-5\sigma$ significance~\cite{Agashe:2020wph}. 
It was further found that the radion channel can be the first discovery channel of KK graviton due to the effectiveness of the cuts designed for the event topology of the radion channel signal, while the KK gluon channel can serve as a cross-check to understand the structure of the underlying model.

\section{Future Prospects} 

We can consider a couple of follow-up studies.  
While a series of our works have shown that the (HL-)LHC would be able to explore parameter space of the extended warped models, the high-energy LHC ($\sqrt{s}=27$~TeV), the 100~TeV future collider 
and a muon collider 
will be excellent venues to explore deep in the parameter space (toward larger masses) or to carry out the precision studies for the models (upon discovery). So, 
it would be interesting (and rather straightforward) to investigate prospects of KK gauge particles and KK graviton at these future colliders, in terms of their experimental reaches and post-discovery analyses. 
Moreover, since such signals are already on the radar of CMS, we hope that the LHC collaborations will pursue this for Run III of the LHC and 
HL-LHC as well.

\begin{figure}[t]
\centering
\includegraphics[width=0.5\linewidth]{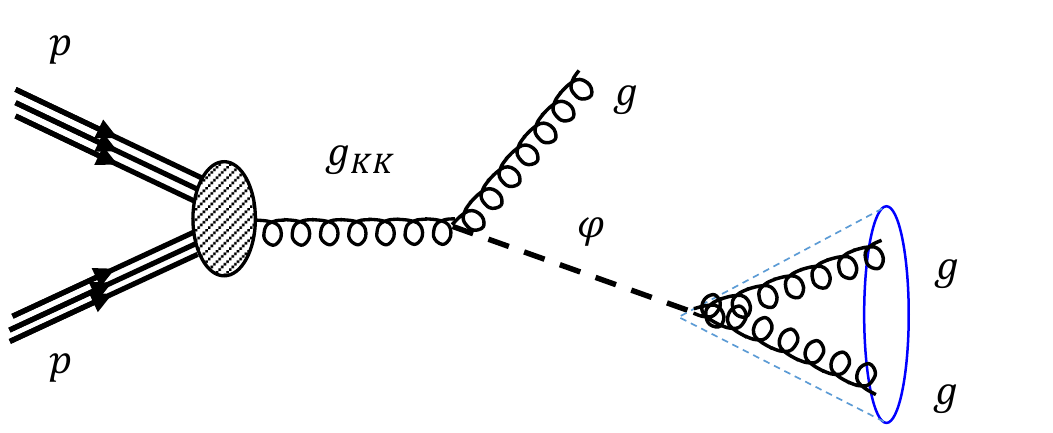}
\caption{Cascade decay of KK gluon into boosted/merged di-gluon, plus an isolated gluon.
Note that this corresponds to the same Feynman diagram as in \autoref{figure:KK-Radion-all}, but with the radion being light, thus boosted.
}
\label{figure:KKg-digluon}
\end{figure}

Second, the above-described KK graviton signals with light radions ($\lesssim 500$~GeV), which are relatively less constrained by existing bounds, are interesting to investigate. 
Once again, such a light radion is allowed for only gluon living in the extended bulk, since radion couplings to SM photons, hence that decay channel is now highly suppressed, even if production via SM gluon fusion is still at the vanilla/full strength.
Owing to substantial mass gaps between such radion and KK states, the gluonic radion (i.e., $\varphi\to gg$) is significantly boosted, hence manifests itself as a merged radion jet (akin to boosted/merged $WW$ mentioned earlier). So, the KK graviton signal in the radion channel will appear as a dijet event, while that in the KK gluon channel will appear as a trijet event. 
Dedicated jet substructure techniques can reveal these signals, allowing us to explore different regions of parameter space.
In fact, CMS has already studied signals for such a light radion arising from decay of {\em direct} production of KK gluon (vs.~from KK graviton mentioned above), giving an isolated gluon, plus a boosted/merged di-gluon (see \autoref{figure:KKg-digluon})
\cite{CMS:2022tqn}.
Obviously, this experimental jet substructure algorithm developed for the boosted/merged di-gluon can also be used for the KK {\em graviton}
case in \autoref{figure:KK-graviton-event-topology}.

The challenging and universal nature of the class of signals discussed in this white paper is useful for providing a well-motivated testing ground for novel data driven methods. For example, Ref.~\cite{Filipek:2021qbe} used machine learning (ML) techniques to identify such signals, i.e., with 
BSM resonances decaying into ``fat'' object made of SM particles, thus with a multi-prong structure. Further, this topology was an important and one of the most challenging ones to detect in anomaly detection challenges~\cite{Kasieczka:2021xcg}, which inform the community of the limitations of existing methods, and allow development of new ones.

\section{Conclusions} 
Models of extended warped extra dimensions are an attractive framework as they retain the virtue of addressing the gauge/flavor hierarchy problems like the standard ones {\em and} contain new particles of a few TeV (or even lighter), allowing the models to be tested even in the upcoming LHC runs. 
A summary of the generalized tri-boson-type signals from gauge KK covering all possibilities is shown in \autoref{table:triboson}, which includes cases where the two SM gauge bosons from radion decay are merged.
A similar compilation for ``quadri''-boson signals from KK graviton can be easily done, generalizing \autoref{figure:KK-graviton-event-topology} to the case with {\em all} SM gauge fields residing in the extended bulk.
Furthermore, collider techniques for searching for such new particles can inspire the search effort for other new physics models giving similar experimental signatures.  
In particular, the studies above serves as motivation to analyse cascade decays of BSM resonances {\em in general}, i.e., within other frameworks/models as well: again, instead of directly decaying into SM particles, BSM heavy particle can readily decay into other/lighter
BSM particles which subsequently decay into SM particles.
Given rich phenomenology and potential impacts on collider physics, the research on the extended warped model will be an important aspect of not only the energy-frontier program but the theory-frontier program in the next decade.  

\begin{table}[t]
\centering
{
\begin{tabular}{ >{\centering\arraybackslash} m{0.19\textwidth} || >{\centering\arraybackslash}m{0.15\textwidth} || >{\centering\arraybackslash}m{0.18\textwidth} | >{\centering\arraybackslash}m{0.18\textwidth} |>{\centering\arraybackslash}m{0.18\textwidth}  }
\hline 
~ & \multicolumn{4}{c}{Singal features}\\
\cline{2-5}
 Radion mass ~~~~~~ (KK scale $\gtrsim 3$ TeV) & General topology &  (I). All SM gauge in the extended bulk \cite{Agashe:2016kfr} & (II). Only EW  in the extended bulk \cite{Agashe:2017wss,Agashe:2018leo}& (III). Only QCD  in the extended bulk \cite{Agashe:2020wph} \\
\hline
\hline
Heavy radion ~~~($\gtrsim 1$ TeV) & isolated three bosons
& { mixture} of gluons and $ W/Z/ \gamma$ (Fig.~\ref{figure:KK-Radion-all})& { mixture} of $ W/Z/ \gamma$ (Fig.~\ref{figure:KK-Radion-EW}) & 3 gluons/jets \\
\hline
Light radion ($O(100)$ GeV)& { boosted} { di}-boson $+$ an isolated boson  & (ruled out by di-photon searches) & $ W / Z / \gamma + $ boosted  $ WW / ZZ / Z \gamma / \gamma \gamma $ (Figs.~\ref{figure:KKW-diW} and \ref{figure:KKphoton-phi-diphoton}) & gluon $+$  boosted di-gluon (Fig.~\ref{figure:KKg-digluon})\\
\hline
\end{tabular}
}
\caption{Summary of various triboson signals from KK gauge particle production and decay into radion.}
\label{table:triboson}
\end{table}

\section*{Acknowledgments} 
The work of KA and DS was supported in part by the NSF grant PHY-1914731 and by the Maryland Center for Fundamental Physics. 
The work of JHC is supported by the U.S. Department of Energy, Office of Science under contract DE-AC02-76SF00515.
PD is supported in part by Simons Investigator in Physics Award 623940 and NSF award PHY-1915093. 
The work of ME was supported by the Swiss National Science Foundation under contract 200020-188671 and through the National Center of Competence in Research SwissMAP.
The work of SH is supported by a DOE grant DE-SC-0013642 and a DOE grant DE-AC02-06CH11357. 
The work of DK is supported by the DOE Grant No. DE-SC0010813. 
The work of RKM is supported by the National Science Foundation under Grant No. NSF PHY-1748958 and NSF PHY-1915071.

\newpage

\bibliography{ref}
\bibliographystyle{JHEP}

\end{document}